\documentclass[letter]{aa}
\usepackage[varg]{txfonts}
\usepackage{graphicx}
\usepackage{natbib}
\usepackage{gensymb}
\usepackage{xcolor}
\usepackage{hyperref}
\usepackage{lineno}
\bibpunct{(}{)}{;}{a}{}{,}

\begin{document}
\nolinenumbers

\title{UGC\,10043 in depth: Dissecting the polar bulge and subtle low surface brightness features}

\author{S. K. H. Bahr\inst{1} \& A. V. Mosenkov\inst{1}}

\institute{Department of Physics and Astronomy, N283 ESC, Brigham Young University, Provo, UT 84602, USA}

\date{Received / Accepted }

\abstract{
Galaxies with polar structures (of which polar-ring galaxies (PRGs) are a prominent subclass) contain components that are kinematically decoupled and highly inclined relative to the major axis of the host galaxy. Modern deep optical surveys provide a powerful means of detecting low surface brightness (LSB) features around galaxies, which offers critical insights into the formation and evolution of galaxies with polar structures.
UGC\,10043 is an edge-on galaxy that is notable for its prominent bulge, which extends orthogonally to the disk plane. In addition, the galaxy displays a well-defined integral-shaped disk warp and multiple dust features crossing the bulge along the minor galaxy axis.
We present new deep optical photometry of UGC\,10043 down to $\mu_g=29.5$~mag\,arcsec$^{-2}$ and perform a detailed analysis of its LSB and polar structures. The observations reveal a stellar stream aligned along the polar axis, alongside other signatures of tidal interaction, including a flat, tilted LSB envelope that extends toward the neighboring galaxy MCG\,+04-37-035, with which UGC\,10043 is connected by an H{\sc i} bridge. Our results suggest that the polar component of UGC\,10043 comprises an older, triaxial polar bulge and a younger, forming polar structure that likely originates from the ongoing disruption of a dwarf satellite galaxy. It also simultaneously participates in active interaction with MCG\,+04-37-035.
}


\maketitle

\section{Introduction} \label{sec:intro}

Galaxies with polar structures (PSGs) are rare and unique systems that are characterized by the orthogonal orientation of the major axes of the host galaxy and its polar component \citep{1990AJ....100.1489W, moiseev11, 2012MNRAS.422.2386F, 2015MNRAS.447.2287R}. Several types of polar structures are known, including polar rings and disks (both inner and outer) \citep{1999ApJ...519L.127B, 2020MNRAS.497.2039M}, polar bulges \citep{2012AstBu..67..147M, 2015AstL...41..748R}, and polar halos \citep{2016ApJ...823...19C, 2020MNRAS.494.1751M, 2021MNRAS.506.5030M}.

In polar-ring galaxies (PRGs), the central host is typically an early-type galaxy that is relatively gas poor, while the polar structure is generally a star-forming ring enriched with subsolar metallicity gas \citep{2010ApJ...714.1081S, 2015A&A...583A..48I, 2019MNRAS.486.4186E}. Polar rings are characterized by a distinct gap in the optical range that separates the polar structure from the host galaxy, whereas polar disks display a continuous distribution of material without an optical gap. Polar rings and disks can both be classified as outer, where the host galaxy is surrounded by a polar structure of a similar or larger size, or inner, where the polar structure is embedded within the larger host galaxy \citep[see e.g.][]{1990AJ....100.1489W, 2012AstBu..67..147M}. 

Polar bulges share similarities with inner polar rings and disks in that the host galaxy is larger than the polar structure. However, unlike inner polar rings or disks, the morphology of polar bulges is similar to that of normal bulges, except for their distinct triaxial, polar orientation: As shown by \citealt{2010MNRAS.401..559M,2012A&AT...27..325S}, traditional bulges in edge-on galaxies align with their stellar disks and display apparent flattening ranging from 0.3 to 0.8. Compared with the three-dimensional shape of polar bulges, inner polar rings and disks are also flatter structures. Polar bulges and inner polar rings and disks typically rotate almost orthogonally to the disk of the host galaxy \citep{2012AstBu..67..147M, 2015AstL...41..748R}. Moreover, preexisting polar bulges can facilitate the formation of inner polar rings and disks, as observed in NGC\,4698 \citep{2012MNRAS.423L..79C}.

\citet{2015AstL...41..748R} note that the morphological distinction between polar-bulge and polar-ring or disk galaxies is somewhat subjective because it primarily depends on the relative sizes and brightnesses of their components. The same study demonstrated, however, that polar-bulge galaxies exhibit bulge S\'ersic indices and bulge-disk color differences characteristic of late-type spiral galaxies. For this study, we define polar-bulge galaxies as spiral or S0 hosts with bulges that, while typical in most respects, are distinguished by their elongation in the polar direction.

The currently proposed formation mechanisms of PSGs include major mergers \citep{hibbard95, bekki97, bekki98}, minor mergers \citep{johnston01, 2011MNRAS.414.3645S}, tidal accretion \citep{reshetnikovsotnikova97, bournaudcombes03}, and intergalactic gas infall along cosmological filaments \citep{thakarryden96, thakarryden98, macchio06, brook08}. Each of these scenarios has been confirmed by simulations and agrees with one or more observational findings \citep[see e.g.][]{reshetnikov05, stanonik09, ordenesbriceno16, 2022RAA....22k5003M}. This highlights that PSGs are valuable probes for studying diverse galaxy formation and evolution processes.

Deep optical surveys ($\sim$29--30~mag\,arcsec$^{-2}$) enable the identification and characterization of low surface brightness (LSB) polar structures that remain undetectable at the depths achieved by most standard surveys (e.g., the Sloan Digital Sky Survey, SDSS, which reaches a depth of 26.5~mag\,arcsec$^{-2}$ in the $r$ band). By uncovering previously unknown LSB features, deep photometric observations provide crucial insights into galactic morphology that are inaccessible with shallower data \citep{2015MNRAS.446..120D, 2022MNRAS.514.4898V, 2023A&A...671A.141M}. For instance, while the occurrence rate of PSGs was previously estimated at only 0.1\% because the resolution of existing optical surveys was limited, recent deep imaging suggests that the true occurrence rate may be as high as 1--3\% \citep{2023MNRAS.525.4663D, 2024A&A...681L..15M}. 

Deep imaging has been used to expand the sample of candidate PRGs \citep{2022RAA....22k5003M}, but similar efforts have not yet been applied to other types of polar structures, such as polar bulges. We use deep optical imaging to investigate LSB structures associated with the polar-bulge galaxy UGC\,10043. Our findings demonstrate that deep imaging of PSGs can provide crucial insights into the formation mechanisms of their polar structures, and consequently, into the evolutionary history of their host galaxies. Thus, deep imaging will be essential for future studies that explore the structure, formation, environment, and broader characteristics of PSGs. By treating PSGs as a coherent class of related objects and not as isolated phenomena, we can identify the underlying physical patterns that govern these unique systems.

UGC\,10043 is an edge-on Sbc spiral galaxy located at a distance of 31.46~Mpc \citep{2021ApJ...914..104B}. Following the prescriptions of \citet{1998AJ....115.2264T}, \citet{2004AJ....128..137M}, and \citet{2005ApJS..160..149S}, we calculated a {\it B}-band luminosity of $1.1 \times 10^8\,L_{\bigodot}$ and a dynamical mass of $9.9 \times 10^{10}\,M_{\bigodot}$ for UGC\,10043. It is a candidate polar-bulge system \citep{2015AstL...41..748R} and displays several distinctive features. These include a superthin disk \citep{2021ApJ...914..104B}, a prominent major-axis dust lane that cleanly divides the galaxy into two symmetrical halves, emphasizing its pure edge-on orientation, a triaxial bulge elongated along the minor axis, and a narrow dust lane aligned with the minor axis \citep{2004AJ....128..137M, 2019MNRAS.488..590B}. Its complex morphology, coupled with a large-scale galactic wind driven by a central starburst \citep{aguirre09}, suggests a history that involved at least one secondary event. 

UGC\,10043 is also a member of the small galaxy group [TSK2008] 1238 \citep{2008ApJ...676..184T, 2013AJ....146...86T}, in which tidal interactions with the neighboring galaxy MCG\,+04-37-035 (confirmed by the presence of an H{\sc i} bridge between the two) likely influenced its structure \citep{aguirre09}. We analyze deep optical observations of UGC\,10043 that reach a surface brightness limit of 29.5~mag\,arcsec$^{-2}$. They reveal a complex set of previously undetected LSB structures for the first time. These findings provide new insights into the formation of UGC\,10043 and offer a broader perspective on the evolutionary pathways of galaxies with polar bulges.

This paper is organized as follows. In Section~\ref{sec:data} we describe the data collection and reduction process for the deep photometric images. Section~\ref{sec:results} describes our image analysis. The implications of the results are discussed in Section~\ref{sec:discuss}, and we summarize our conclusions in Section~\ref{sec:conc}.

We adopt a standard flat $\Lambda$CDM cosmology with $\Omega_{m}$ = 0.308, $\Omega_{\Lambda}$ = 0.692, and $H_0=67.8$ km\,s$^{-1}$\,Mpc$^{-1}$. All magnitudes are reported in the AB photometric system.

\section{Data} \label{sec:data}

Images of UGC\,10043 were obtained from the Dark Energy Spectroscopic Instrument (DESI) Legacy Survey DR10 \citep{2019AJ....157..168D} in the $g$ and $r$ bands, which reach photometric depths of 29.2 and 28.5~mag\,arcsec$^{-2}$, respectively\footnote{The photometric depth is measured throughout this work within a 10\textquotesingle\textquotesingle\,$\times$\,10\textquotesingle\textquotesingle\, box at the $3\sigma$ level.}. In addition, we acquired new deep optical observations using the ARCTIC imager on the 3.5-meter Astrophysical Research Consortium (ARC) telescope at Apache Point Observatory, which achieves depths of 29.44~mag\,arcsec$^{-2}$ ($g$ band) and 29.38~mag\,arcsec$^{-2}$ ($r$ band). Each filter was observed with five 15-minute exposures, yielding average PSF FWHM values of 0.67\textquotesingle\textquotesingle\, in $g$ and 0.75\textquotesingle\textquotesingle\, in $r$. These deep optical images were primarily used to analyze LSB features in the outer regions of the galaxy. We followed the method of \citet{2022RAA....22k5003M} in preparing the deep photometric data using the Python package IMage ANalysis (IMAN) \citep{2020MNRAS.497.2039M}, including initial data reduction (bias and dark correction, flat-field correction, cosmic-ray removal, and astrometric and photometric calibration) for the ARC 3.5 m data and rigorous masking and background subtraction for all images.

While deep optical images are better suited for analyzing low LSB features, the inner structure of the galaxy is more effectively traced in the near-infrared, where the effects of dust extinction are significantly reduced, despite some contribution from PAH emission. This allows for a clearer view of the underlying stellar structure, in particular, along the galaxy plane and in central regions that are highly obscured in the optical. For this purpose, we used an IRAC 3.6~$\mu$m image from the {\it Spitzer} S$^4$G survey \citep{2010PASP..122.1397S}, which offers a plate scale of 0.75\textquotesingle\textquotesingle/pixel and a PSF FWHM of 1.7\textquotesingle\textquotesingle. Although the photometric depth of the IRAC image is shallower (27~mag\,arcsec$^{-2}$), it is more than sufficient for modeling the stellar structure of this galaxy. 

\begin{figure}
    \centering
    \includegraphics[width=0.45\textwidth]{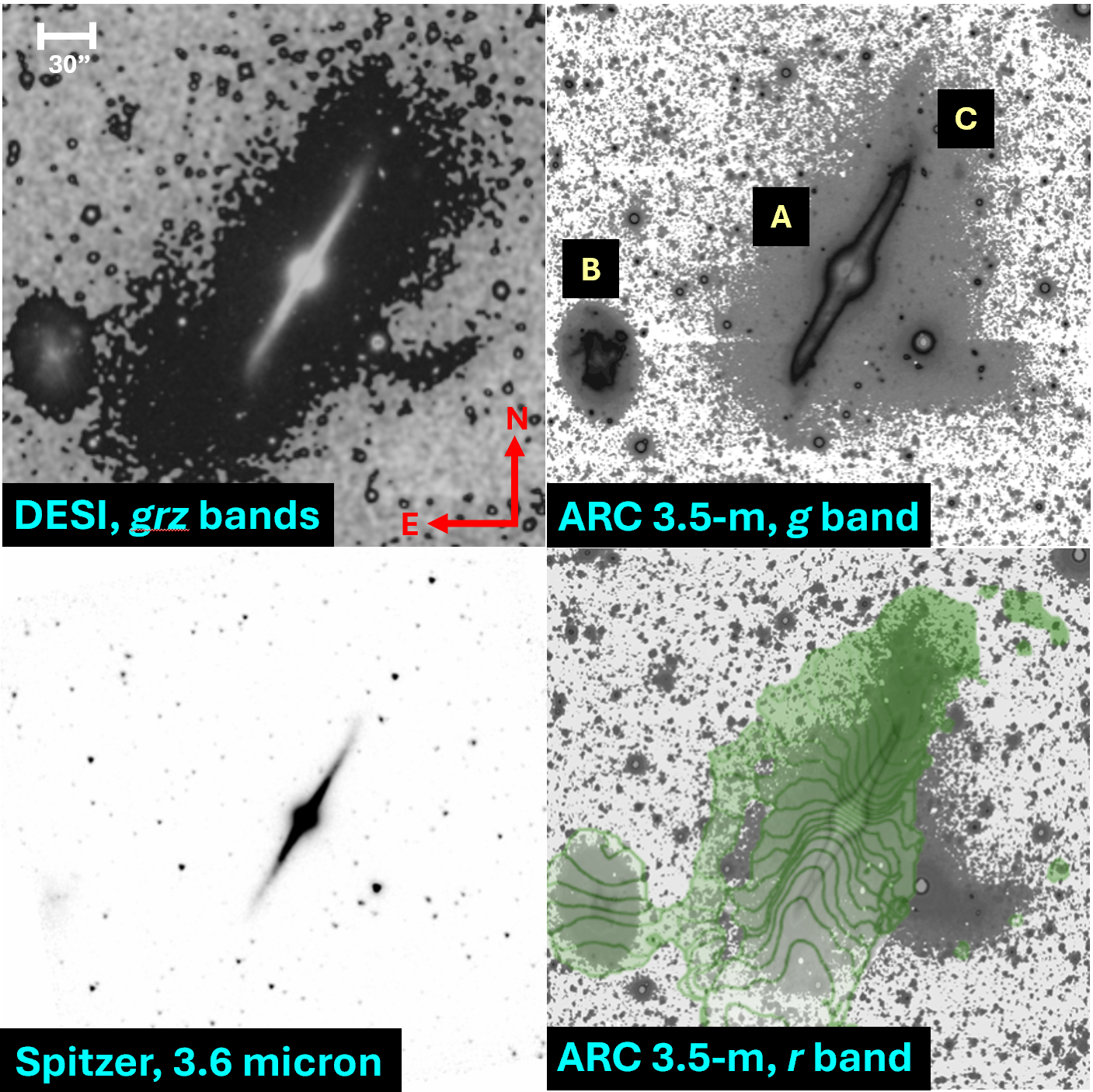}
    \caption{Four views of UGC\,10043 (clockwise from top left): The DESI Legacy Survey, the ARC 3.5 m telescope {\it g} and {\it r} bands, and the {\it Spitzer} Space Telescope. The ARC 3.5 m and DESI Legacy images have been enhanced to highlight the LSB stream and warp. The positions of UGC\,10043, MCG\,+04-37-035, and MD2004 dwarf C are indicated in the ARC 3.5 m {\it g}-band image by the letters A, B, and C, respectively. The H{\sc i} map from \citet{aguirre09} is overlaid on the ARC 3.5 m {\it r}-band image.}
    \label{fig:10043_images}
\end{figure}

\section{Results}\label{sec:results}

\subsection{Low-surface brightness features}\label{sec:lsb}

\begin{figure}
    \centering
    \includegraphics[width=0.8\linewidth]{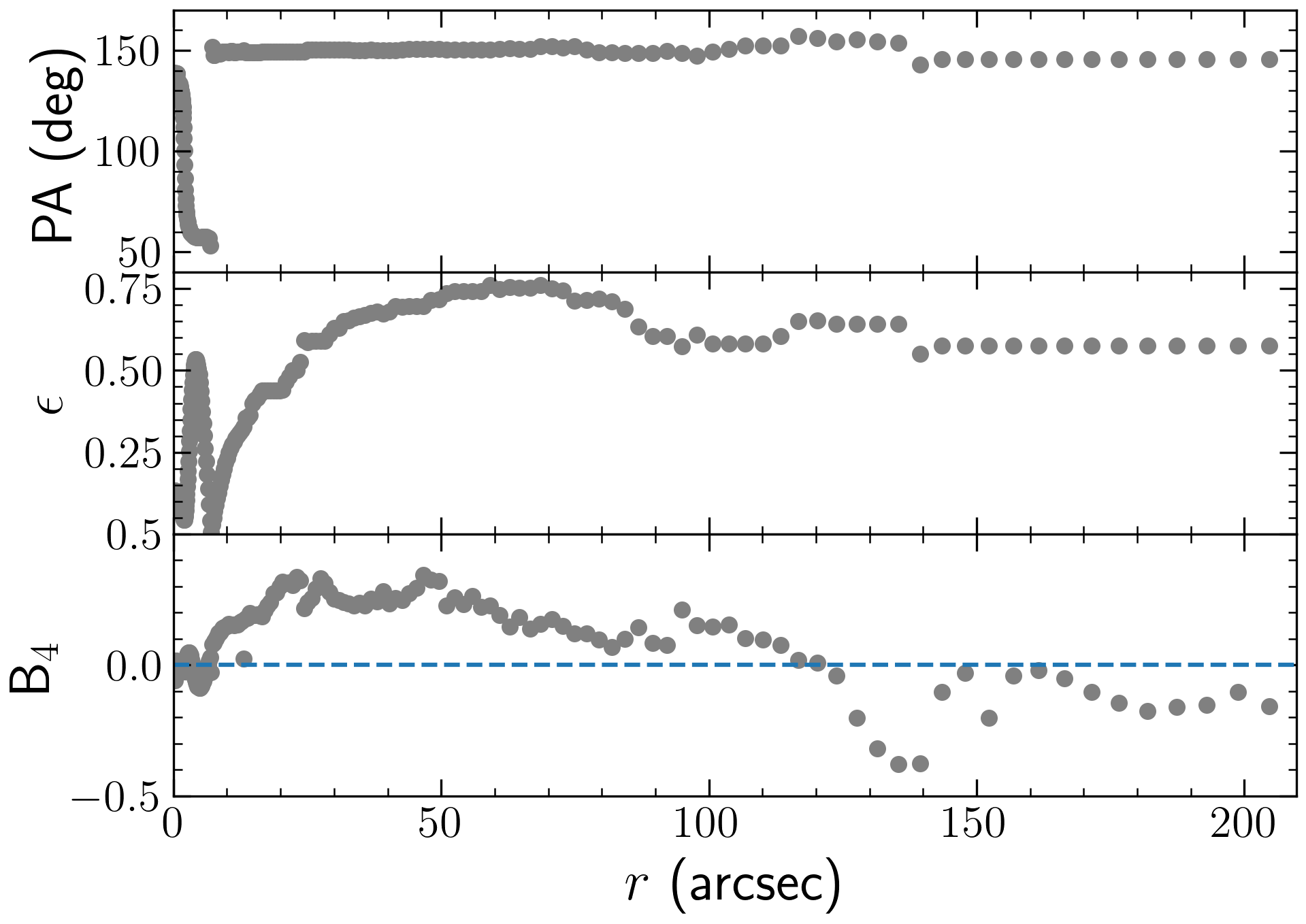}
    \caption{Isophote fitting results: Position angle, ellipticity, and $B4$ for UGC\,10043 as a function of distance from the center of the galaxy.}
    \label{fig:isophote}
\end{figure}

The DESI and ARC deep photometries, as shown in Fig.~\ref{fig:10043_images}, reveal a variety of low surface brightness (LSB) features surrounding UGC\,10043. To quantify the galaxy isophotes, we performed isophote fitting using IRAF/STSDAS on the DESI Legacy {\it grz}-band image. The results are presented in Fig.~\ref{fig:isophote} and highlight significant variations in position angle, ellipticity, and the B4 Fourier mode, which characterizes the shape of the isophotes (disky if $B4>0$, and boxy if $B4<0$). In the central region, the isophotes appear to be approximately elliptical, with a noticeable change in isophote parameters due to the presence of the polar structure. Beyond a radius of 10\textquotesingle\textquotesingle, the stellar disk becomes dominant and exhibits disky isophotes. Farther out, beyond 125\textquotesingle\textquotesingle, the tilted thick envelope displays boxy isophotes, which indicate another structural transition in the galaxy. Notably, the optical envelope is tilted by $6.5\degree$ from the plane of the galaxy, as indicated by the shift in position angle around 140\textquotesingle\textquotesingle\, in Fig.~\ref{fig:isophote}, and it appears to link UGC\,10043 to MCG\,+04-37-035 (see the DESI Legacy image, Fig.~\ref{fig:10043_images}). However, this tilt is in the opposite direction compared to the disk warp (described below) and the gaseous envelope reported by \citet{aguirre09}, as shown in Fig.~\ref{fig:10043_images}. The outermost isophote of this boxy, tilted envelope has a surface brightness of 28.8~mag\,arcsec$^{-2}$, with a semimajor axis of 33.0~kpc and a semiminor axis of 19.3~kpc.

The integral-sign warp in the outer regions of the disk is well documented in the literature. Our deep photometry reveals, however, that this warp has a rather irregular shape, as seen in our deep ARC images (Fig.~\ref{fig:10043_images}). The warp angle is measured as the inclination between the galaxy plane and the line connecting the tip of the outermost isophote. It is $11.4\degree$ for the northern warp and $11.3\degree$ for the southern warp at the isophote level of 31~mag\,arcsec$^{-2}$. The warp begins at a distance of 7.14~kpc for the northern warp and 6.26~kpc for the southern warp.

One of the most striking LSB features of UGC\,10043 is the stellar stream that extends southwest of the galaxy. This stream emerges from the galactic bulge at an angle of $82.9\degree$ from the galactic plane. It almost coincides with the alignment of the polar bulge. 
The stream has a distinct curved shape. Following the curve of the stream in our ARC 3.5-m {\it r}-band image, we measured a total projected stream length of 10.9~kpc at the isophote level 31~mag\,arcsec$^{-2}$. The stream also appears to loop around the galaxy on the northeast side of the disk.

\subsection{Photometric model} \label{sec:models}

The photometric decomposition of UGC\,10043 was performed to explore its structure and obtain the parameters of the disk and bulge, including an additional polar bulge component. A photometric model for UGC\,10043 was created using IMAN to interact with {\sc GALFIT} \citep{2010AJ....139.2097P}. No LSB features are visible in the {\it Spitzer} image and therefore do not affect our decomposition. We describe the creation and results of the best-fit model below.

To determine the appropriate components and initial parameters for the photometric decomposition, several preliminary {\sc GALFIT} runs were conducted. The finalized photometric model consisted of two edge-on isothermal disks that represent the thin and thick disks, each with a truncation, and of two S\'ersic functions to account for the bulge and the polar bulge. To assess the robustness of the final model and to estimate confidence intervals for the parameter values, we performed 1000 {\sc GALFIT} runs. In each run, the best-fit parameters obtained previously were used as the initial guess, and we randomly masked 10\% of the pixels (bootstrapping). The fitting procedure employed the {\it Spitzer} PSF and the weight images provided in the S$^4$G database.

\begin{figure}
    \centering
    \includegraphics[width=0.45\textwidth]{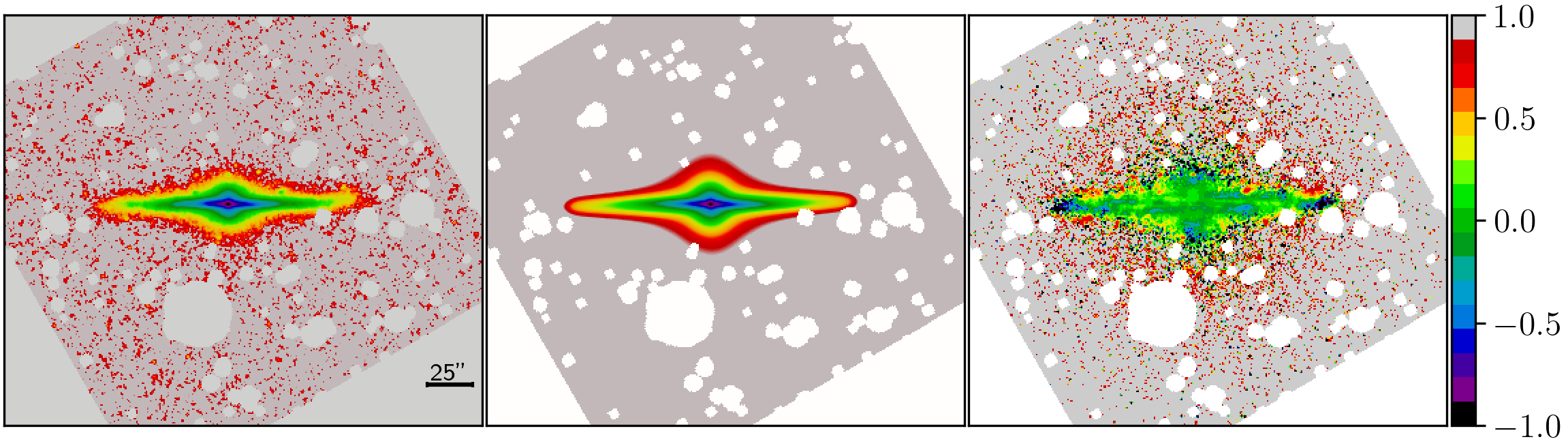}
    \caption{Best-fit results for {\it Spitzer} data. Left to right: Observed surface brightness, best-fit model surface brightness, and relative residuals.}
    \label{fig:10043_fit_spi}
\end{figure}

Table~\ref{10043fitvals_spi} reports the mean best-fit parameter values and the $1\sigma$ confidence intervals from these 1000 runs. All parameters have small errors, except for the position angle of the polar bulge, which nonetheless supports the interpretation that the secondary central component is indeed polar. The observed and model surface brightness maps, along with the residuals, are shown in Fig.~\ref{fig:10043_fit_spi}. The model accurately reproduces the galaxy structure, as indicated by the green regions in the residuals. 

\begin{table}
\caption{Best-fit model parameters for the {\it Spitzer} data.}
\label{10043fitvals_spi}
\centering
\begin{tabular}{|c c|} 
 \hline
 Parameter & Value\\
 \hline\hline
 $h_\mathrm{R,thin}$ & $1.07 \pm 0.01$\\
 \hline
 $z_\mathrm{0,thin}$ & $0.08 \pm 0.01$\\
 \hline
 $PA_\mathrm{thin}$ & -31 (fixed)\\
 \hline
 $\mu_\mathrm{e,thin}$ & $16.26 \pm 0.03$\\
 \hline
 $h_\mathrm{R,thick}$ & $3.18 \pm 0.02$\\
 \hline
 $z_\mathrm{0,thick}$ & $0.42 \pm 0.01$\\
 \hline
 $PA_\mathrm{thick}$ & $-29.07 \pm 0.02$\\
 \hline
 $\mu_\mathrm{e,thick}$ & $19.51 \pm 0.02$\\
 \hline
  $n_\mathrm{bulge}$ & $0.40 \pm 0.05$\\
 \hline
 $r_\mathrm{e,bulge}$ & $1.13 \pm 0.06$\\
 \hline
 $\mu_\mathrm{e,bulge}$ & $20.83 \pm 0.10$\\
 \hline
 $b/a_\mathrm{bulge}$ & $0.82 \pm 0.04$\\
 \hline
 $n_\mathrm{polar\,bulge}$ & $3.79 \pm 0.30$\\
 \hline
 $r_\mathrm{e,polar\,bulge}$ & $0.52 \pm 0.03$\\
 \hline
 $\mu_\mathrm{e,polar\,bulge}$ & $19.36 \pm 0.05$\\
 \hline
 $b/a_\mathrm{polar\,bulge}$ & $0.61 \pm 0.02$\\
 \hline
 $PA_\mathrm{polar\,bulge}$ & $58.50 \pm 1.25$\\
 \hline
\end{tabular}
\tablefoot{Scale lengths $h_\mathrm{R}$ and scale heights $z_0$ (both in kpc), effective radius $r_\mathrm{e}$ (kpc), position angle $PA$ (degrees), effective surface brightness $\mu_\mathrm{e}$ (mag\,arcsec$^{-2}$), S\'ersic index $n$, and axis ratio $b/a$.}
\end{table}

\section{Discussion} \label{sec:discuss}

We confirm that UGC\,10043 hosts a polar bulge with an inclination angle of $89.5\degree$ relative to the galactic disk. Our decomposition of the central component into two bulge components provides further evidence that UGC\,10043 is a polar-bulge galaxy. The {\it Spitzer} photometry reveals a flattened ($b/a=0.61$), brighter ($\mu_e=19.36$~mag\,arcsec$^{-2}$) polar bulge with a relatively high S\'ersic index of $3.7$, characteristic of classical bulges \citep{2008AJ....136..773F}. This polar bulge coexists with a rounder ($b/a=0.82$), dimmer ($\mu_e=20.83$~mag\,arcsec$^{-2}$) bulge with an unusually low S\'ersic index of $0.4$. 

\citet{2004AJ....128..137M} found that the bulge of UGC\,10043 is triaxial, with prolate inner isophotes elongated perpendicular to the disk and outer isophotes that twist to become oblate and nearly circular. Although the authors did not detect compelling evidence for orthogonal rotation, the observed isophotal structure and the presence of a minor-axis dust lane suggest a component with misaligned angular momentum. This behavior is unusual for typical Sbc galaxies, but consistent with polar-bulge systems \citep{1999ApJ...519L.127B,2000A&A...360..439S,2002A&A...383..390R,2003A&A...408..873C, 2012AstBu..67..147M, 2015AstL...41..748R}. According to \citet{2022MNRAS.512.2537G}, a S\'ersic index $\ge 2$ in spiral galaxies from the IllustrisTNG50 simulation is associated with a higher fraction of ex situ stars. Thus, the S\'ersic index of the polar bulge uncovered in our decomposition, combined with the morphological and tentative kinematic evidence from \citet{2004AJ....128..137M}, supports the interpretation that the combined bulge in UGC\,10043 is a structure that formed through at least one secondary event, such as tidal accretion or a minor merger.

The difference in position angle between the thin and thick disk in UGC\,10043 is $1.9 \pm 0.02$ degrees, and the outer optical envelope is tilted by $6.5\degree$ compared to the main galaxy disk. These values are consistent with the results reported for tilted edge-on disk galaxies by \citet{2020MNRAS.497.2039M} ($5\pm2.3\degree$ and $3.1\pm0.64\degree$ for NGC\,4452 and NGC\,4469, respectively). The magnitude of the change in tilt with increasingly deep isophotes is also typical of the galaxies reported by \citet{2020MNRAS.497.2039M}, but the outer envelope of UGC\,10043 is tilted in the opposite direction compared to the shallower isophotes. 

Numerous theories have been proposed to explain the formation and prevalence of galactic warps. One widely discussed mechanism involves galaxy interactions and satellite accretion \citep[e.g.,][]{1997ApJ...480..503H, 2014ApJ...789...90K, 2020MNRAS.498.3535S}. Other proposed explanations include the influence of misaligned dark matter halos \citep{1995ApJ...442..492D}, discrete bending modes within a self-gravitating disk \citep{1988MNRAS.234..873S}, interactions between the gaseous disk and extragalactic magnetic fields \citep{1990A&A...236....1B}, and cosmic infall onto a disk galaxy \citep{2006MNRAS.370....2S}. Observational evidence suggests, however, that the most prominent warps are typically found in denser environments \citep[e.g.,][]{1998A&A...337....9R, 1999A&AS..138..101R, 2001A&A...373..402S, 2006NewA...11..293A, 2016JKAS...49..239A}. 

Observational evidence suggests that the structure of UGC\,10043 has evolved through interactions with other galaxies. Located within a group environment, UGC\,10043 is physically connected to its neighboring galaxy MCG\,+04-37-035, as shown by both H{\sc i} observations \citep{aguirre09} and optical imaging (see Fig.~\ref{fig:10043_images}). This connection indicates that MCG\,+04-37-035 likely contributed gas (and possibly stars) to the polar structure of UGC\,10043 through tidal accretion, given the linked envelopes of the two galaxies.

A possible dwarf spheroidal galaxy, MD2004 dwarf C, is also located near the far edge of the disk. It may be a companion to UGC\,10043, and potentially, a third progenitor of the southwestern stream. However, MD2004 dwarf C is an unlikely source of gas for either the stream or the polar bulge, as its position is within the tilted LSB envelope, but poorly aligned with the gas flow. Therefore, the more plausible scenario is that the gas (and possibly stellar material) originated from MCG\,+04-37-035. The ongoing central starburst in UGC\,10043, likely fueled by tidally accreted gas, powers the galactic wind, while the star formation rate in the remainder of the galaxy remains relatively low \citep{2017MNRAS.467.4951L}. Furthermore, the previously detected LSB stream \citep{2011A&A...536A..66M} that we confirmed in this study likely represents the tidal disruption of a dwarf satellite galaxy. This reinforces the evidence that the complex history of UGC\,10043 included multiple interactions.

\section{Conclusions} \label{sec:conc}

From the set of polar bulge galaxies identified by \citet{2015AstL...41..748R}, we have selected a case study that exhibits evidence of recent or ongoing tidal interactions. The goal of this study was to investigate tidal interaction and accretion as potential mechanisms for the formation and extended lifetime of polar structures. We conducted a structural analysis of UGC\,10043 using {\it Spitzer} 3.6~$\mu$m data. This resulted in a best-fit model that included a thin and thick disk, a Gaussian bulge, and a polar structure. We also used deep optical photometry of UGC\,10043, including DESI DR10 observations and our own ARC 3.5 m imaging in the $g$ and $r$ wavebands, to examine LSB features that surround the galaxy.

Our main conclusions are listed below.

\begin{itemize}
    \item The {\it Spitzer} 3.6~$\mu$m data reveal a thick disk that is tilted by almost 2 degrees relative to the thin disk.  
    \item Our photometric decomposition of the {\it Spitzer} image also indicates a Gaussian ($n=0.4$) and a polar ($n=3.7$) bulge (inclination angle 89.5 degrees). 
    \item The thick disk transitions into a thick envelope with a vertical extent of 19.3 kpc, showing an even more pronounced tilt of 6.5 degrees from the thin disk for the outermost isophotes. The radius of the envelope reaches 33.0 kpc.  
    \item Deep optical observations confirmed that UGC\,10043 is connected to the neighboring galaxy MCG\,+04-37-035, consistent with previous H{\sc i} observations from \citet{aguirre09}.  
    \item \citet{2011A&A...536A..66M} presented an enhanced SDSS DR7 image of UGC\,10043 showing the southwestern LSB stream (see their Fig.~18), but did not discuss this feature. We confirmed the presence of this stream using deeper images, and we demonstrated that it extends 10.9~kpc in the polar direction, with a projected inclination angle of $82.9\degree$ from the galactic plane.   
    \item These LSB features around UGC\,10043 suggest possible formation mechanisms for the polar structure and the pronounced disk warp of the galaxy that likely involved multiple accretion processes. These processes may include interactions with MCG\,+04-37-035 and a potential dwarf galaxy that caused the southwestern LSB stream. It is also plausible that other distinctive characteristics of this galaxy, such as its ongoing starburst, can also be attributed to tidal accretion. 
\end{itemize}

To further investigate the role of tidal interactions in the formation and evolution of diverse polar structures, it will be essential to compare these results with in-depth studies of other galaxies exhibiting various polar structure subtypes, and with PSGs from cosmological simulations. Future work will require deep optical observations of a large sample of PSGs to gain a comprehensive understanding of their formation mechanisms. The identification of the most common processes that cause polar structures will provide crucial insights into how PSGs form and evolve, and how they fit within the broader hierarchical model of galaxy evolution.

\begin{acknowledgements}
We thank the anonymous referee for their helpful comments that improved this paper.
 
Based in part on observations obtained with the Apache Point Observatory 3.5-meter telescope, which is owned and operated by the Astrophysical Research Consortium.

The Legacy Surveys consist of three individual and complementary projects: the Dark Energy Camera Legacy Survey (DECaLS; NOAO Proposal ID \# 2014B-0404; PIs: David Schlegel and Arjun Dey), the Beijing-Arizona Sky Survey (BASS; NOAO Proposal ID \# 2015A-0801; PIs: Zhou Xu and Xiaohui Fan), and the Mayall z-band Legacy Survey (MzLS; NOAO Proposal ID \# 2016A-0453; PI: Arjun Dey). DECaLS, BASS and MzLS together include data obtained, respectively, at the Blanco telescope, Cerro Tololo Inter-American Observatory, National Optical Astronomy Observatory (NOAO); the Bok telescope, Steward Observatory, University of Arizona; and the Mayall telescope, Kitt Peak National Observatory, NOAO. The Legacy Surveys project is honored to be permitted to conduct astronomical research on Iolkam Du’ag (Kitt Peak), a mountain with particular significance to the Tohono O’odham Nation.

Some of the data presented in this paper were obtained from the Mikulski Archive for Space Telescopes (MAST). STScI is operated by the Association of Universities for Research in Astronomy, Inc., under NASA contract NAS5-26555. Support for MAST for non-HST data is provided by the NASA Office of Space Science via grant NNX13AC07G and by other grants and contracts. 

This research made use of NASA’s Astrophysics Data System for bibliographic information.

This research has made use of the NASA/IPAC Extragalactic Database (NED) and the NASA/IPAC Infrared Science Archive, which are funded by the National Aeronautics and Space Administration and operated by the California Institute of Technology.
\end{acknowledgements}

\bibliographystyle{aa}
\bibliography{bibliography}

\begin{thebibliography}{67}
\expandafter\ifx\csname natexlab\endcsname\relax\def\natexlab#1{#1}\fi

\bibitem[{{Aguirre} {et~al.}(2009){Aguirre}, {Uson}, \& {Matthews}}]{aguirre09}
{Aguirre}, P., {Uson}, J.~M., \& {Matthews}, L.~D. 2009, in Revista Mexicana de
  Astronomia y Astrofisica Conference Series, Vol.~35, Revista Mexicana de
  Astronomia y Astrofisica Conference Series, 201--202

\bibitem[{{Ann} \& {Bae}(2016)}]{2016JKAS...49..239A}
{Ann}, H.~B. \& {Bae}, H.~J. 2016, Journal of Korean Astronomical Society, 49,
  239

\bibitem[{{Ann} \& {Park}(2006)}]{2006NewA...11..293A}
{Ann}, H.~B. \& {Park}, J.~C. 2006, \na, 11, 293

\bibitem[{{Battaner} {et~al.}(1990){Battaner}, {Florido}, \&
  {Sanchez-Saavedra}}]{1990A&A...236....1B}
{Battaner}, E., {Florido}, E., \& {Sanchez-Saavedra}, M.~L. 1990, \aap, 236, 1

\bibitem[{{Bekki}(1997)}]{bekki97}
{Bekki}, K. 1997, \apjl, 490, L37

\bibitem[{{Bekki}(1998)}]{bekki98}
{Bekki}, K. 1998, \apj, 499, 635

\bibitem[{{Bertola} {et~al.}(1999){Bertola}, {Corsini}, {Beltr{\'a}n},
  {Pizzella}, {Sarzi}, {Cappellari}, \& {Funes}}]{1999ApJ...519L.127B}
{Bertola}, F., {Corsini}, E.~M., {Beltr{\'a}n}, J.~C.~V., {et~al.} 1999, \apjl,
  519, L127

\bibitem[{{Bizyaev} {et~al.}(2021){Bizyaev}, {Makarov}, {Reshetnikov},
  {Mosenkov}, {Kautsch}, \& {Antipova}}]{2021ApJ...914..104B}
{Bizyaev}, D., {Makarov}, D.~I., {Reshetnikov}, V.~P., {et~al.} 2021, \apj,
  914, 104

\bibitem[{{Bournaud} \& {Combes}(2003)}]{bournaudcombes03}
{Bournaud}, F. \& {Combes}, F. 2003, \aap, 401, 817

\bibitem[{{Brook} {et~al.}(2008){Brook}, {Governato}, {Quinn}, {Wadsley},
  {Brooks}, {Willman}, {Stilp}, \& {Jonsson}}]{brook08}
{Brook}, C.~B., {Governato}, F., {Quinn}, T., {et~al.} 2008, \apj, 689, 678

\bibitem[{{Buta}(2019)}]{2019MNRAS.488..590B}
{Buta}, R.~J. 2019, \mnras, 488, 590

\bibitem[{{Corsini} {et~al.}(2012){Corsini}, {M{\'e}ndez-Abreu}, {Pastorello},
  {Dalla Bont{\`a}}, {Morelli}, {Beifiori}, {Pizzella}, \&
  {Bertola}}]{2012MNRAS.423L..79C}
{Corsini}, E.~M., {M{\'e}ndez-Abreu}, J., {Pastorello}, N., {et~al.} 2012,
  \mnras, 423, L79

\bibitem[{{Corsini} {et~al.}(2003){Corsini}, {Pizzella}, {Coccato}, \&
  {Bertola}}]{2003A&A...408..873C}
{Corsini}, E.~M., {Pizzella}, A., {Coccato}, L., \& {Bertola}, F. 2003, \aap,
  408, 873

\bibitem[{{Crnojevi{\'c}} {et~al.}(2016){Crnojevi{\'c}}, {Sand}, {Spekkens},
  {Caldwell}, {Guhathakurta}, {McLeod}, {Seth}, {Simon}, {Strader}, \&
  {Toloba}}]{2016ApJ...823...19C}
{Crnojevi{\'c}}, D., {Sand}, D.~J., {Spekkens}, K., {et~al.} 2016, \apj, 823,
  19

\bibitem[{{Deg} {et~al.}(2023){Deg}, {Palleske}, {Spekkens}, {Wang}, {Jarrett},
  {English}, {Lin}, {Yeung}, {Mould}, {Catinella}, {D{\'e}nes}, {Elagali},
  {For}, {Kamphuis}, {Koribalski}, {Lee-Waddell}, {Murugeshan}, {Oh}, {Rhee},
  {Serra}, {Westmeier}, {Wong}, {Bekki}, {Bosma}, {Carignan}, {Holwerda}, \&
  {Yu}}]{2023MNRAS.525.4663D}
{Deg}, N., {Palleske}, R., {Spekkens}, K., {et~al.} 2023, \mnras, 525, 4663

\bibitem[{{Dey} {et~al.}(2019){Dey}, {Schlegel}, {Lang}, {Blum}, {Burleigh},
  {Fan}, {Findlay}, {Finkbeiner}, {Herrera}, {Juneau}, {Landriau}, {Levi},
  {McGreer}, {Meisner}, {Myers}, {Moustakas}, {Nugent}, {Patej}, {Schlafly},
  {Walker}, {Valdes}, {Weaver}, {Y{\`e}che}, {Zou}, {Zhou}, {Abareshi},
  {Abbott}, {Abolfathi}, {Aguilera}, {Alam}, {Allen}, {Alvarez}, {Annis},
  {Ansarinejad}, {Aubert}, {Beechert}, {Bell}, {BenZvi}, {Beutler}, {Bielby},
  {Bolton}, {Brice{\~n}o}, {Buckley-Geer}, {Butler}, {Calamida}, {Carlberg},
  {Carter}, {Casas}, {Castander}, {Choi}, {Comparat}, {Cukanovaite}, {Delubac},
  {DeVries}, {Dey}, {Dhungana}, {Dickinson}, {Ding}, {Donaldson}, {Duan},
  {Duckworth}, {Eftekharzadeh}, {Eisenstein}, {Etourneau}, {Fagrelius},
  {Farihi}, {Fitzpatrick}, {Font-Ribera}, {Fulmer}, {G{\"a}nsicke},
  {Gaztanaga}, {George}, {Gerdes}, {Gontcho}, {Gorgoni}, {Green}, {Guy},
  {Harmer}, {Hernandez}, {Honscheid}, {Huang}, {James}, {Jannuzi}, {Jiang},
  {Joyce}, {Karcher}, {Karkar}, {Kehoe}, {Kneib}, {Kueter-Young}, {Lan},
  {Lauer}, {Le Guillou}, {Le Van Suu}, {Lee}, {Lesser}, {Perreault Levasseur},
  {Li}, {Mann}, {Marshall}, {Mart{\'\i}nez-V{\'a}zquez}, {Martini}, {du Mas des
  Bourboux}, {McManus}, {Meier}, {M{\'e}nard}, {Metcalfe},
  {Mu{\~n}oz-Guti{\'e}rrez}, {Najita}, {Napier}, {Narayan}, {Newman}, {Nie},
  {Nord}, {Norman}, {Olsen}, {Paat}, {Palanque-Delabrouille}, {Peng},
  {Poppett}, {Poremba}, {Prakash}, {Rabinowitz}, {Raichoor}, {Rezaie},
  {Robertson}, {Roe}, {Ross}, {Ross}, {Rudnick}, {Safonova}, {Saha},
  {S{\'a}nchez}, {Savary}, {Schweiker}, {Scott}, {Seo}, {Shan}, {Silva},
  {Slepian}, {Soto}, {Sprayberry}, {Staten}, {Stillman}, {Stupak}, {Summers},
  {Sien Tie}, {Tirado}, {Vargas-Maga{\~n}a}, {Vivas}, {Wechsler}, {Williams},
  {Yang}, {Yang}, {Yapici}, {Zaritsky}, {Zenteno}, {Zhang}, {Zhang}, {Zhou}, \&
  {Zhou}}]{2019AJ....157..168D}
{Dey}, A., {Schlegel}, D.~J., {Lang}, D., {et~al.} 2019, \aj, 157, 168

\bibitem[{{Dubinski} \& {Kuijken}(1995)}]{1995ApJ...442..492D}
{Dubinski}, J. \& {Kuijken}, K. 1995, \apj, 442, 492

\bibitem[{{Duc} {et~al.}(2015){Duc}, {Cuillandre}, {Karabal}, {Cappellari},
  {Alatalo}, {Blitz}, {Bournaud}, {Bureau}, {Crocker}, {Davies}, {Davis}, {de
  Zeeuw}, {Emsellem}, {Khochfar}, {Krajnovi{\'c}}, {Kuntschner}, {McDermid},
  {Michel-Dansac}, {Morganti}, {Naab}, {Oosterloo}, {Paudel}, {Sarzi}, {Scott},
  {Serra}, {Weijmans}, \& {Young}}]{2015MNRAS.446..120D}
{Duc}, P.-A., {Cuillandre}, J.-C., {Karabal}, E., {et~al.} 2015, \mnras, 446,
  120

\bibitem[{{Egorov} \& {Moiseev}(2019)}]{2019MNRAS.486.4186E}
{Egorov}, O.~V. \& {Moiseev}, A.~V. 2019, \mnras, 486, 4186

\bibitem[{{Finkelman} {et~al.}(2012){Finkelman}, {Funes}, \&
  {Brosch}}]{2012MNRAS.422.2386F}
{Finkelman}, I., {Funes}, J.~G., \& {Brosch}, N. 2012, \mnras, 422, 2386

\bibitem[{{Fisher} \& {Drory}(2008)}]{2008AJ....136..773F}
{Fisher}, D.~B. \& {Drory}, N. 2008, \aj, 136, 773

\bibitem[{{Gargiulo} {et~al.}(2022){Gargiulo}, {Monachesi}, {G{\'o}mez},
  {Nelson}, {Pillepich}, {Pakmor}, {Grand}, {Fragkoudi}, {Hernquist}, {Lovell},
  \& {Marinacci}}]{2022MNRAS.512.2537G}
{Gargiulo}, I.~D., {Monachesi}, A., {G{\'o}mez}, F.~A., {et~al.} 2022, \mnras,
  512, 2537

\bibitem[{{Hibbard} \& {Mihos}(1995)}]{hibbard95}
{Hibbard}, J.~E. \& {Mihos}, J.~C. 1995, \aj, 110, 140

\bibitem[{{Huang} \& {Carlberg}(1997)}]{1997ApJ...480..503H}
{Huang}, S. \& {Carlberg}, R.~G. 1997, \apj, 480, 503

\bibitem[{{Iodice} {et~al.}(2015){Iodice}, {Coccato}, {Combes}, {de Zeeuw},
  {Arnaboldi}, {Weilbacher}, {Bacon}, {Kuntschner}, \&
  {Spavone}}]{2015A&A...583A..48I}
{Iodice}, E., {Coccato}, L., {Combes}, F., {et~al.} 2015, \aap, 583, A48

\bibitem[{{Johnston} {et~al.}(2001){Johnston}, {Sackett}, \&
  {Bullock}}]{johnston01}
{Johnston}, K.~V., {Sackett}, P.~D., \& {Bullock}, J.~S. 2001, \apj, 557, 137

\bibitem[{{Kim} {et~al.}(2014){Kim}, {Peirani}, {Kim}, {Ann}, {An}, \&
  {Yoon}}]{2014ApJ...789...90K}
{Kim}, J.~H., {Peirani}, S., {Kim}, S., {et~al.} 2014, \apj, 789, 90

\bibitem[{{L{\'o}pez-Cob{\'a}} {et~al.}(2017){L{\'o}pez-Cob{\'a}},
  {S{\'a}nchez}, {Moiseev}, {Oparin}, {Bitsakis}, {Cruz-Gonz{\'a}lez},
  {Morisset}, {Galbany}, {Bland-Hawthorn}, {Roth}, {Dettmar}, {Bomans},
  {Gonz{\'a}lez Delgado}, {Cano-D{\'\i}az}, {Marino}, {Kehrig}, {Monreal
  Ibero}, \& {Abril-Melgarejo}}]{2017MNRAS.467.4951L}
{L{\'o}pez-Cob{\'a}}, C., {S{\'a}nchez}, S.~F., {Moiseev}, A.~V., {et~al.}
  2017, \mnras, 467, 4951

\bibitem[{{Macci{\`o}} {et~al.}(2006){Macci{\`o}}, {Moore}, \&
  {Stadel}}]{macchio06}
{Macci{\`o}}, A.~V., {Moore}, B., \& {Stadel}, J. 2006, \apjl, 636, L25

\bibitem[{{Mart{\'\i}nez-Delgado} {et~al.}(2023){Mart{\'\i}nez-Delgado},
  {Cooper}, {Rom{\'a}n}, {Pillepich}, {Erkal}, {Pearson}, {Moustakas},
  {Laporte}, {Laine}, {Akhlaghi}, {Lang}, {Makarov}, {Borlaff}, {Donatiello},
  {Pearson}, {Mir{\'o}-Carretero}, {Cuillandre}, {Dom{\'\i}nguez},
  {Roca-F{\`a}brega}, {Frenk}, {Schmidt}, {G{\'o}mez-Flechoso}, {Guzman},
  {Libeskind}, {Dey}, {Weaver}, {Schlegel}, {Myers}, \&
  {Valdes}}]{2023A&A...671A.141M}
{Mart{\'\i}nez-Delgado}, D., {Cooper}, A.~P., {Rom{\'a}n}, J., {et~al.} 2023,
  \aap, 671, A141

\bibitem[{{Mart{\'\i}nez-Delgado} {et~al.}(2021){Mart{\'\i}nez-Delgado},
  {Rom{\'a}n}, {Erkal}, {Schirmer}, {Roca-F{\`a}brega}, {Laine}, {Donatiello},
  {Jimenez}, {Malin}, \& {Carballo-Bello}}]{2021MNRAS.506.5030M}
{Mart{\'\i}nez-Delgado}, D., {Rom{\'a}n}, J., {Erkal}, D., {et~al.} 2021,
  \mnras, 506, 5030

\bibitem[{{Matthews} \& {de Grijs}(2004)}]{2004AJ....128..137M}
{Matthews}, L.~D. \& {de Grijs}, R. 2004, \aj, 128, 137

\bibitem[{{Miskolczi} {et~al.}(2011){Miskolczi}, {Bomans}, \&
  {Dettmar}}]{2011A&A...536A..66M}
{Miskolczi}, A., {Bomans}, D.~J., \& {Dettmar}, R.~J. 2011, \aap, 536, A66

\bibitem[{{Moiseev}(2012)}]{2012AstBu..67..147M}
{Moiseev}, A.~V. 2012, Astrophysical Bulletin, 67, 147

\bibitem[{{Moiseev} {et~al.}(2011){Moiseev}, {Smirnova}, {Smirnova}, \&
  {Reshetnikov}}]{moiseev11}
{Moiseev}, A.~V., {Smirnova}, K.~I., {Smirnova}, A.~A., \& {Reshetnikov}, V.~P.
  2011, \mnras, 418, 244

\bibitem[{{Mosenkov} {et~al.}(2020{\natexlab{a}}){Mosenkov}, {Rich}, {Koch},
  {Brosch}, {Thilker}, {Rom{\'a}n}, {M{\"u}ller}, {Smirnov}, \&
  {Usachev}}]{2020MNRAS.494.1751M}
{Mosenkov}, A., {Rich}, R.~M., {Koch}, A., {et~al.} 2020{\natexlab{a}}, \mnras,
  494, 1751

\bibitem[{{Mosenkov} {et~al.}(2024){Mosenkov}, {Bahr}, {Reshetnikov},
  {Shakespear}, \& {Smirnov}}]{2024A&A...681L..15M}
{Mosenkov}, A.~V., {Bahr}, S.~K.~H., {Reshetnikov}, V.~P., {Shakespear}, Z., \&
  {Smirnov}, D.~V. 2024, \aap, 681, L15

\bibitem[{{Mosenkov} {et~al.}(2022){Mosenkov}, {Reshetnikov}, {Skryabina}, \&
  {Shakespear}}]{2022RAA....22k5003M}
{Mosenkov}, A.~V., {Reshetnikov}, V.~P., {Skryabina}, M.~N., \& {Shakespear},
  Z. 2022, Research in Astronomy and Astrophysics, 22, 115003

\bibitem[{{Mosenkov} {et~al.}(2020{\natexlab{b}}){Mosenkov}, {Smirnov},
  {Sil'chenko}, {Rich}, {Reshetnikov}, \& {Kormendy}}]{2020MNRAS.497.2039M}
{Mosenkov}, A.~V., {Smirnov}, A.~A., {Sil'chenko}, O.~K., {et~al.}
  2020{\natexlab{b}}, \mnras, 497, 2039

\bibitem[{{Mosenkov} {et~al.}(2010){Mosenkov}, {Sotnikova}, \&
  {Reshetnikov}}]{2010MNRAS.401..559M}
{Mosenkov}, A.~V., {Sotnikova}, N.~Y., \& {Reshetnikov}, V.~P. 2010, \mnras,
  401, 559

\bibitem[{{Ordenes-Brice{\~n}o} {et~al.}(2016){Ordenes-Brice{\~n}o},
  {Georgiev}, {Puzia}, {Goudfrooij}, \& {Arnaboldi}}]{ordenesbriceno16}
{Ordenes-Brice{\~n}o}, Y., {Georgiev}, I.~Y., {Puzia}, T.~H., {Goudfrooij}, P.,
  \& {Arnaboldi}, M. 2016, \aap, 585, A156

\bibitem[{{Peng} {et~al.}(2010){Peng}, {Ho}, {Impey}, \&
  {Rix}}]{2010AJ....139.2097P}
{Peng}, C.~Y., {Ho}, L.~C., {Impey}, C.~D., \& {Rix}, H.-W. 2010, \aj, 139,
  2097

\bibitem[{{Reshetnikov} {et~al.}(2005){Reshetnikov}, {Bournaud}, {Combes},
  {Fa{\'u}ndez-Abans}, {de Oliveira-Abans}, {van Driel}, \&
  {Schneider}}]{reshetnikov05}
{Reshetnikov}, V., {Bournaud}, F., {Combes}, F., {et~al.} 2005, \aap, 431, 503

\bibitem[{{Reshetnikov} \& {Combes}(1998)}]{1998A&A...337....9R}
{Reshetnikov}, V. \& {Combes}, F. 1998, \aap, 337, 9

\bibitem[{{Reshetnikov} \& {Combes}(1999)}]{1999A&AS..138..101R}
{Reshetnikov}, V. \& {Combes}, F. 1999, \aaps, 138, 101

\bibitem[{{Reshetnikov} \& {Combes}(2015)}]{2015MNRAS.447.2287R}
{Reshetnikov}, V. \& {Combes}, F. 2015, \mnras, 447, 2287

\bibitem[{{Reshetnikov} \& {Sotnikova}(1997)}]{reshetnikovsotnikova97}
{Reshetnikov}, V. \& {Sotnikova}, N. 1997, \aap, 325, 933

\bibitem[{{Reshetnikov} {et~al.}(2002){Reshetnikov}, {Fa{\'u}ndez-Abans}, \&
  {de Oliveira-Abans}}]{2002A&A...383..390R}
{Reshetnikov}, V.~P., {Fa{\'u}ndez-Abans}, M., \& {de Oliveira-Abans}, M. 2002,
  \aap, 383, 390

\bibitem[{{Reshetnikov} {et~al.}(2015){Reshetnikov}, {Savchenko}, {Mosenkov},
  {Sotnikova}, \& {Bizyaev}}]{2015AstL...41..748R}
{Reshetnikov}, V.~P., {Savchenko}, S.~S., {Mosenkov}, A.~V., {Sotnikova},
  N.~Y., \& {Bizyaev}, D.~V. 2015, Astronomy Letters, 41, 748

\bibitem[{{Sarzi} {et~al.}(2000){Sarzi}, {Corsini}, {Pizzella}, {Vega
  Beltr{\'a}n}, {Cappellari}, {Funes}, \& {Bertola}}]{2000A&A...360..439S}
{Sarzi}, M., {Corsini}, E.~M., {Pizzella}, A., {et~al.} 2000, \aap, 360, 439

\bibitem[{{Schwarzkopf} \& {Dettmar}(2001)}]{2001A&A...373..402S}
{Schwarzkopf}, U. \& {Dettmar}, R.~J. 2001, \aap, 373, 402

\bibitem[{{Semczuk} {et~al.}(2020){Semczuk}, {{\L}okas}, {D'Onghia},
  {Athanassoula}, {Debattista}, \& {Hernquist}}]{2020MNRAS.498.3535S}
{Semczuk}, M., {{\L}okas}, E.~L., {D'Onghia}, E., {et~al.} 2020, \mnras, 498,
  3535

\bibitem[{{Shen} \& {Sellwood}(2006)}]{2006MNRAS.370....2S}
{Shen}, J. \& {Sellwood}, J.~A. 2006, \mnras, 370, 2

\bibitem[{{Sheth} {et~al.}(2010){Sheth}, {Regan}, {Hinz}, {Gil de Paz},
  {Men{\'e}ndez-Delmestre}, {Mu{\~n}oz-Mateos}, {Seibert}, {Kim},
  {Laurikainen}, {Salo}, {Gadotti}, {Laine}, {Mizusawa}, {Armus},
  {Athanassoula}, {Bosma}, {Buta}, {Capak}, {Jarrett}, {Elmegreen},
  {Elmegreen}, {Knapen}, {Koda}, {Helou}, {Ho}, {Madore}, {Masters},
  {Mobasher}, {Ogle}, {Peng}, {Schinnerer}, {Surace}, {Zaritsky},
  {Comer{\'o}n}, {de Swardt}, {Meidt}, {Kasliwal}, \&
  {Aravena}}]{2010PASP..122.1397S}
{Sheth}, K., {Regan}, M., {Hinz}, J.~L., {et~al.} 2010, \pasp, 122, 1397

\bibitem[{{Sil'chenko} {et~al.}(2011){Sil'chenko}, {Chilingarian}, {Sotnikova},
  \& {Afanasiev}}]{2011MNRAS.414.3645S}
{Sil'chenko}, O.~K., {Chilingarian}, I.~V., {Sotnikova}, N.~Y., \& {Afanasiev},
  V.~L. 2011, \mnras, 414, 3645

\bibitem[{{Sotnikova} {et~al.}(2012){Sotnikova}, {Reshetnikov}, \&
  {Mosenkov}}]{2012A&AT...27..325S}
{Sotnikova}, N.~Y., {Reshetnikov}, V.~P., \& {Mosenkov}, A.~V. 2012,
  Astronomical and Astrophysical Transactions, 27, 325

\bibitem[{{Sparke} \& {Casertano}(1988)}]{1988MNRAS.234..873S}
{Sparke}, L.~S. \& {Casertano}, S. 1988, \mnras, 234, 873

\bibitem[{{Spavone} {et~al.}(2010){Spavone}, {Iodice}, {Arnaboldi}, {Gerhard},
  {Saglia}, \& {Longo}}]{2010ApJ...714.1081S}
{Spavone}, M., {Iodice}, E., {Arnaboldi}, M., {et~al.} 2010, \apj, 714, 1081

\bibitem[{{Springob} {et~al.}(2005){Springob}, {Haynes}, {Giovanelli}, \&
  {Kent}}]{2005ApJS..160..149S}
{Springob}, C.~M., {Haynes}, M.~P., {Giovanelli}, R., \& {Kent}, B.~R. 2005,
  \apjs, 160, 149

\bibitem[{{Stanonik} {et~al.}(2009){Stanonik}, {Platen}, {Arag{\'o}n-Calvo},
  {van Gorkom}, {van de Weygaert}, {van der Hulst}, \& {Peebles}}]{stanonik09}
{Stanonik}, K., {Platen}, E., {Arag{\'o}n-Calvo}, M.~A., {et~al.} 2009, \apjl,
  696, L6

\bibitem[{{Thakar} \& {Ryden}(1996)}]{thakarryden96}
{Thakar}, A.~R. \& {Ryden}, B.~S. 1996, \apj, 461, 55

\bibitem[{{Thakar} \& {Ryden}(1998)}]{thakarryden98}
{Thakar}, A.~R. \& {Ryden}, B.~S. 1998, \apj, 506, 93

\bibitem[{{Tully} {et~al.}(2013){Tully}, {Courtois}, {Dolphin}, {Fisher},
  {H{\'e}raudeau}, {Jacobs}, {Karachentsev}, {Makarov}, {Makarova},
  {Mitronova}, {Rizzi}, {Shaya}, {Sorce}, \& {Wu}}]{2013AJ....146...86T}
{Tully}, R.~B., {Courtois}, H.~M., {Dolphin}, A.~E., {et~al.} 2013, \aj, 146,
  86

\bibitem[{{Tully} {et~al.}(1998){Tully}, {Pierce}, {Huang}, {Saunders},
  {Verheijen}, \& {Witchalls}}]{1998AJ....115.2264T}
{Tully}, R.~B., {Pierce}, M.~J., {Huang}, J.-S., {et~al.} 1998, \aj, 115, 2264

\bibitem[{{Tully} {et~al.}(2008){Tully}, {Shaya}, {Karachentsev}, {Courtois},
  {Kocevski}, {Rizzi}, \& {Peel}}]{2008ApJ...676..184T}
{Tully}, R.~B., {Shaya}, E.~J., {Karachentsev}, I.~D., {et~al.} 2008, \apj,
  676, 184

\bibitem[{{Vera-Casanova} {et~al.}(2022){Vera-Casanova}, {G{\'o}mez},
  {Monachesi}, {Gargiulo}, {Pallero}, {Grand}, {Marinacci}, {Pakmor},
  {Simpson}, {Frenk}, \& {Morales}}]{2022MNRAS.514.4898V}
{Vera-Casanova}, A., {G{\'o}mez}, F.~A., {Monachesi}, A., {et~al.} 2022,
  \mnras, 514, 4898

\bibitem[{{Whitmore} {et~al.}(1990){Whitmore}, {Lucas}, {McElroy},
  {Steiman-Cameron}, {Sackett}, \& {Olling}}]{1990AJ....100.1489W}
{Whitmore}, B.~C., {Lucas}, R.~A., {McElroy}, D.~B., {et~al.} 1990, \aj, 100,
  1489

\end{thebibliography}

\end{document}